\documentclass[a4paper,10pt]{article}
\usepackage{ucs}
\usepackage[utf8]{inputenc}
\usepackage{amsmath}
\usepackage{amsfonts}
\usepackage{amssymb}
\usepackage{amsthm}
\usepackage{rotating}
\usepackage{xcolor}
\usepackage[english]{babel}
\usepackage[T1]{fontenc}
\usepackage{graphicx}
\usepackage{hyperref}
\usepackage{cuted}
\usepackage{verbatim}
\usepackage{subfigure}
\usepackage{multirow}
\usepackage{bbold}
\usepackage{slashed}
\usepackage{indentfirst}
\usepackage{tcolorbox}
\usepackage{placeins}
\usepackage[a4paper, total={7in, 10in}]{geometry}
\usepackage{setspace}

\newcommand{\bpsi}{\bar{\psi}}

\begin{document}

\title{ \bf  Pion constituent quarks  couplings strong form factors: 
 a dynamical approach 
}

\author{ F\'abio L. Braghin
\\
{\normalsize Instituto de F\'\i sica, Federal University of Goias,
 Av. Esperan\c ca, s/n,
 74690-900, Goi\^ania, GO, Brazil}
}

\date{}

\maketitle

\begin{abstract} 
Form factors for pions interactions with constituent quarks are investigated
as the leading 
effective couplings obtained from
 a one loop background field method  applied to a global color model.
 Two pion field definitions are considered and  the 
resulting eleven 
 form factors are expressed 
 in terms of
components of the quark and gluon propagators
that compose only two momentum dependent
functions. 
A
 momentum dependent Goldberger Treiman relation is also obtained
as one of the ratios between the form factors.
The resulting form factors with  pion momenta  up 
to 1.5 GeV are exhibitted
for different quark
effective masses and  two different nonperturbative
 gluon propagators and
 they present similar behavior to 
fittings of  experimental data from nucleons form factors. 
The corresponding  pseudoscalar averaged 
quadratic  radii (a.q.r.) and correction to the axial a.q.r.
 are presented
as functions of the sea  quark  effective 
 mass,
 being equal respectively  to the
scalar and vector ones at the present level of calculation.
 \end{abstract}

\section{Introduction}
\label{intro}

Strong,
 electromagnetic and weak
content of hadrons have been   under continous
 intense  theoretical and experimental 
scrutiny.
Different  hadrons   form factors 
are  among the 
main observables for understanding details of their interactions and structures, 
including  sizes, and they are important quantities
to  compare  theoretical  and experimental results 
\cite{exp-r2A,gaillard-savage,choi-etal-93,bardin-etal-1981,andreev-etal-2007}. 
For example,  the vector form factors provide the charge and electromagnetic
hadron structure and interactions,  the nucleon
axial form factor provides important
 information   for  their spin structure and weak interaction
observables such as neutron beta decay or CKM matrix unitarity.
There are many theoretical calculations for the 
light hadrons strong form factors, for example
\cite{beise,maris-craig,charge-pi-rad,drechsel-walcher,yamazaki-etal,constantinou,alexandrou-etal,weise-vogl,hoferichter-etal,ramalho-tsushima-PRD94,eichmann-fischer,exp-bernard+E+meissner,bratt-etal} 
and references therein.
Lattely, lattice estimations  for pion-
nucleon/baryons  interactions were provided
 for progressively 
lower values of the pion mass,
for example in  \cite{bratt-etal,yamazaki-etal,alexandrou-etal}.
 Concerning their very low momentum behavior,
experimental results for  nucleon electromagnetic and strong 
averaged radii provide values
  $\sqrt{<r^2>} \simeq 0.8-0.9$fm \cite{proton-radius,PDG,exp-r2A,hoferichter-etal}.

In spite of the many 
difficulties to provide  a complete description  of hadrons and their interactions
compatible with experimental data directly from QCD,
in particular  in the low and intermediary energies regimes,
  both effective models and effective theories
have been considered  to understand partial or isolated 
aspects of 
 Strong Interactions.
Among these models 
the  constituent quark models (CQM) 
has shown to  describe many aspects of 
 hadron structure and interactions 
by considering  dressed quark degrees of freedom,
 Dynamical Chiral Symmetry Breaking (DChSB) 
and eventually a pion cloud,
 \cite{weinberg-2010,constituent-1,constituent-2,weise-vogl}.
Within  the constituent quark model  it 
has been argued  that  the zero momentum
limit of  the axial  form factor
should be
 $g_A(0)=3/4$ or $g_A(0)=1$ \cite{weise-vogl,weinberg-2010}.
Also, 
a radius  of the order of $0.2-0.3$fm 
has been estimated for  constituent quarks   \cite{weise-vogl,CQ-size}.
  In the Weinberg's large Nc Effective Field Theory 
(EFT)   constituent quarks and gluons interact with 
pions whose dynamics is ruled by the leading terms of Chiral Perturbation 
Theory (ChPT),
coping with the large Nc expansion  \cite{weinberg-2010}.
In  \cite{EPJA-2016,EPJA-2018}
this EFT has been derived as the leading terms from a large quark and gluon
 effective  masses expansion
for the one loop background field method applied to a global color model
in the vacuum and with  leading couplings to the
electromagnetic field.
It can be expected that,  by comparing the 
strong and electromagnetic nucleon
and light mesons form factors with  those for constituent quarks,
the detailed role and contribution of each internal degree of freedom  for the 
details of hadron  structure and interactions 
might be elucidated clearly. 
Of course,  to   accomplish this program, besides further comparisons between
different theoretical frameworks, 
it is also important  to improve the amount  and precision of 
experimental data.
This means that  the related developments  might
shed  light on the partial or even complete  reliability of  CQM-type models to
describe hadron   interactions in particular energy ranges.
Moreover, these comparisons might make explicit particular effects or mechanisms
present in hadrons structure and interactions
by means of analytical or semi-analytical approaches 
besides 
well established lattice QCD framework.
 Eventually it can be used  to assess or to improve
 field theoretic schemes for an eventual unambigous parameterization of 
the nucleon and  nuclear potentials 
 \cite{swanson-etal}.

In the present  work the  strong 
constituent quark  form factors  associated to the leading
pion couplings to 
constituent quarks   are derived and  investigated.
This method was considered before 
for the zero momentum limit of the corresponding pion-constituent quark 
couplings
\cite{EPJA-2016,EPJA-2018}
 and for 
the light vector mesons momentum dependent couplings to constituent quarks
\cite{PRD-2018a,PRD-2018b}. 
The form factors  are  
obtained from  a large quark and gluon effective masses  expansion
 for the one loop background field applied to a global color model.
 The  background field quark becomes the constituent quark due to 
the one loop calculation in which an internal (non perturbative) gluon line dresses
the (background) quark
This is nearly independent from the dynamical symmetry breaking, 
except for the fact that the same gluon propagator required to yield
DChSB is considered.
This momentum dependent 
constituent quark mass  emerges therefore by means of 
a different mechanism from the usual DChSB.
 This might be in agreement with recent  calculations 
\cite{proton-mass}.
The resulting couplings and form factors therefore correspond to  tree level
pion-constituent quark vertices.
These   pion-constituent quarks
 form factors are 
investigated and 
comparisons with experimental data for pion nucleon are presented.
Furthermore
 four further pion derivative couplings with  scalar and  pseudoscalar 
constituent quark currents
that emerge at the same leading terms of the determinant expansion 
are also  presented. 
They might contribute to the vector and axial channels.
Direct and simple momentum dependent 
and independent  relations between different 
 form factors are also presented.
In particular  one relation  corresponds to 
a generalized momentum dependent 
Goldberger Treiman relation (GTR).
Besides that the corresponding strong  quadratic radii 
of constituent quarks (scalar, pseudoscalar, vector and axial) 
 are also presented as functions of the 
quark effective mass. 
The axial (and vector)  pion coupling presented in 
this work provides a further contribution for 
the corresponding axial  (and vector) form factors and quadratic radii
to those calculated in \cite{PRD-2018a}.
Two pion field definitions
are considered, the Weinberg  pion field, in terms of 
covariant derivatives, and the usual parameterization in terms 
of the operators $U=e^{i \vec{\pi} \cdot  \vec\sigma}$.
The conventional definition in terms of the functions 
$U=e^{i\vec{\pi}\cdot \vec{\tau}}$
provides the welll known pseudoscalar pion coupling that is not found 
in the Weinberg  pion field case.
The isospin non degeneracy of up-down quark masses 
is not considered in this work since it should 
be responsible for smaller (higher order)
 effects.
This work is organized as follows.
In the next Section the steps of the method are  briefly  reminded 
and the large quark effective mass expansion of a
sea  quark determinant is performed.
By keeping the full momentum dependence of the resulting constituent quark - pion 
couplings the corresponding  form fators are presented for the two definitions 
of the pion field in the following section.
 Due to the momentum structure of some of the form factors
it is also convenient to perform a truncation that  provides, latter, 
corresponding  positive averaged quadratic  radii.
 All the eleven form factors, five  for the Weinberg pion field and six 
for the second pion field definition, are written in terms of only two 
momentum dependent functions, denoted $F_1(K,Q)$ and 
$F_2(K,Q)$.
Besides that, 
the momentum dependent constituent quark mass 
correction, $M_3(Q)$ is investigated.
In the following  Section numerical results
are  exhibitted  for different values of 
quark effective mass and  for two very different
gluon propagators: an effective longitudinal confining propagator 
 considered  by Cornwall
\cite{cornwall} 
and a transversal one 
used extensively and successfully 
to provide
hadron observables 
by Tandy and Maris \cite{tandy-maris}.
Some ratios and comparisons of the form factors 
are also presented including the estimation of a 
 momentum dependent Goldberger Treiman relation.
The corresponding contributions for the pseudoscalar and axial
strong    constituent quark  quadratic radii
are also investigated as a functions of the quark effective mass
for the different gluon propagators.
In the last Section a summary    is presented.

\section{ The quark determinant, pions  and constituent quark currents  }
\label{sec:two-Q}

Consider the non perturbative one gluon exchange quark-quark interaction 
as 
one of the leading terms of QCD effective action
whose 
 generating functional is given by
\cite{PRC1,ERV}:
\begin{eqnarray} \label{Seff}  
Z &=& N \int {\cal D}[\bpsi, \psi]
\exp \;  i \int_x  \left[
\bar{\psi} \left( i \slashed{\partial} 
- m \right) \psi 
- \frac{g^2}{2}\int_y j_{\mu}^b (x) 
{\tilde{R}}^{\mu \nu}_{bc}  (x-y) j_{\nu}^{c} (y) 
+ \bpsi J + J^* \psi \right] 
\end{eqnarray}
Where 
  $N$ is the normalization, $J,J^*$ the quark  sources,
 $\int_x$ 
stands for 
$\int d^4 x$,
 and
$a,b...=1,...(N_c^2-1)$ stands 
for color in the adjoint representation being $N_c=3$.
The functional measure for the quark field was written as 
${\cal D}[\bpsi, \psi] = {\cal D}[\bpsi] {\cal D} [\psi]$.
The quark gluon coupling constant is assumed to be  $g$
and the development below is  akin to 
the Rainbow Ladder Schwinger Dyson equation (SDE).
Below  indices 
$i,j,k=0,...(N_f^2-1)$ will be used  for SU(2) isospin  indices and therefore $N_f=2$.
 The quark current mass will be assumed to be equal for u, d quarks.
 The color  quark currents are  given by
$j^{\mu}_a = \bar{\psi} \lambda_a \gamma^{\mu} \psi$,
and the sums in color, flavor and Dirac indices are implicit.
A Landau-type gauge will be considered for 
a non pertubative gluon propagator
 that can be written as
$\tilde{R}^{\mu\nu}_{ab}(x-y) \equiv \tilde{R}^{\mu\nu}_{ab} = \delta_{ab} \left[
 \left( g^{\mu\nu} - \frac{\partial^\mu \partial^\nu}{\partial^2}
\right)   R_T  (x-y)
+ \frac{\partial^\mu \partial^\nu}{\partial^2} R_L (x-y) \right]$,
where the 
transversal and  longitudinal components are
 $R_T(x-y)$ and $ R_L (x-y)$.
This non perturbative gluon kernel 
therefore
incorporates
to some extent the gluonic  non Abelian character 
 with a corrected quark-gluon coupling such that they
will provide enough strength to yield dynamical chiral symmetry breaking (DChSB).
This has been found in several   approaches and extensions
\cite{SD-rainbow,maris-craig,kondo,cornwall,higa,holdom,wang-etal}.

The method was  explained  in details in Refs. \cite{PRD-2016,EPJA-2016,EPJA-2018,PRD-2018a,PRD-2018b}
and therefore it will be succintly described below.
A Fierz transformation for the model (\ref{Seff})
 is performed and,  by picking up the 
leading color singlet terms that provide the usual pion couplings,
 it allows to investigate the flavor structure
in a more complete way. Besides that, color singlets,  in one hand, 
avoid problems with unconfined spurious color degrees of freedom
and, on the other hand, provides a direct relation with quark-antiquark 
lightest observed  states.
These states are to be identified with 
 the light hadrons 
degrees of freedom and the scalar chiral condensate 
by means of the corresponding fields to be introduced.
Chiral structures with combinations of  bilocal  currents are  obtained.
The  quark field must be responsible for the formation of mesons and baryons 
and these  different possibilities are envisaged by considering the 
Background Field Method (BFM)
\cite{background,SWbook}.
Therefore  we consider  the  quark field is splitted  into  
sea  quark, $\psi_2$, composing (light)  quark-antiquark states
including light mesons 
and the chiral condensate,
 and the   (constituent)   background quark, 
$\psi_1$, to compose  baryons.
The shift  of   quark bilinears corresponds to performing a one loop  BFM calculation
and it might be written
for each of the color singlet Dirac/isospin 
channels $m=s,p,si,pi,ps,v,a, as,vs$ 
(scalar, pseudoscalar, scalar-isospin triplet, 
pseudoscalar-isospin triplet, vector,  axial, vector-isospin triplet, axial-isospin triplet,
where the isospin singlet states were omitted).
Each of these channels might have a corresponding  auxiliary field.
However only the lightest pseudoscalar-iso-triplet and isoscalar-scalar degrees of 
freedom will be investigated in the present work.
The quark field shift is of the following form:
\begin{eqnarray}
j^m = \bpsi \Gamma^m \psi \to
 (\bpsi \Gamma^m \psi)_2 + (\bpsi \Gamma^m \psi)_1.
\end{eqnarray}
This separation 
   preserves chiral symmetry.
The sea quark can be integrated out exactly by means of 
the auxiliary field method  that give rise to  colorless quark-antiquark states, 
light mesons and the chiral quark condensate.
Auxilary fields are introduced by means of 
the   unity integrals multiplying the generating functional.
The only  degrees of freedom considered in this work 
are the chiral scalar and pseudoscalar- iso-triplet ones 
which are needed for the pion sector in the leading order.
The heavier vector and axial mesons can be neglected in the 
lower energy regime.
Therefore one will be  left with a model 
for pions and a scalar field interacting wtih constituent quarks.
The corresponding   unity integral for the scalar and pseudoscalar
 auxiliary 
bilocal fields $S(x,y), P_i(x,y)$ is the following:
\begin{eqnarray}
 1 &=& N'' \int D[S] D[P_i]
 e^{- \frac{i}{2 }  
\int_{x,y}   R (x-y)  \alpha  \left[ (S - g   j^S_{(2)})^2 +
(P_i -  g    j^{P}_{i,(2)} )^2 \right]}
,
\end{eqnarray}
where $N''$ is a normalization, and 
\begin{eqnarray} \label{gluonpro-R}
R(x-y) = 3 R_T (-y) + R_L (x-y).
\end{eqnarray}
Bilocal auxiliary fields for the  different flavors  
 can be expanded in an infinite orthogonal basis with 
all the excitations in the corresponding channel.
For the pseudoscalar isotriplet fields one has:\begin{eqnarray}
P_i (x,y) = P_i  \left( \frac{x+y}{2}, x-y \right)
= P_i (u ,z) =  
 \sum_k F_{k} (z) P_{i,k} (u),
\end{eqnarray}
where $ F_{k}$ are vacuum functions invariant under translation for
 each of the   local field $P^\mu_{i,k} (u)$.
For the low energy regime one  might
 pick up only the lowest energy modes, lighest $k=0$
which corresponds to the pions in this channel, i.e. $P_{i,k=0} = \pi_i$,
 making the form factors  
to reduce to constants in the zero momentum  limit
$F_{k}(z) = F_{k} (0)$.
The saddle point equations for each of the remaining
auxiliary fields, after the integration of the sea quark,
  can be written from the condition:
$\frac{\partial S_{eff}}{\partial \phi_q} = 0$.
These equations for the NJL model and for the model (\ref{Seff})
with Schwinger Dyson equations at the rainbow ladder level
  have been analyzed in many works
in the vacuum or under a  finite energy density.
The scalar field has the only saddle point equation
with non trivial solution for the quark-antiquark chiral condensate.
This classical solution generates an effective mass for 
sea quarks.
 
Chiral symmetry leaves a freedom to define the pion field and chiral 
rotations can be done to modify its definition.
The scalar field   can be frozen  by means of a 
chiral rotation and this produces the chiral condensate and 
a strongly non linear pion sector.
An usual  pion field definition  is parameterized by
the functions: $U= exp(i \vec{\pi} \cdot \vec{\sigma})$ and 
$U^\dagger= exp( - i \vec{\pi} \cdot \vec{\sigma})$.
 To investigate  this aspect
another  pion field definition,
the Weinberg ones,  is   characterized by writing 
all the chiral invariant sector in terms of
a covariant pion derivative given by:
\begin{eqnarray}
{\cal D}_\mu {\pi}_i = \frac{\partial_\mu \pi_i}{1 + \vec{\pi}^2}.
\end{eqnarray}
The chiral symmetry breaking  terms however can  depend on 
combinations of $\vec{\pi}$ and $\vec{\pi}^2$.
By doing the corresponding chiral rotations particular set of
constituent quark-pion interactions are obtained.
 The corresponding Jacobian of the path integral measure will not be calculated
and it 
 might induce extra  terms for the resulting form factors. 

By  performing a Gaussian integration of the sea  quark field,
the resulting determinant can be written, 
by means of the identity
$\det A = \exp \; Tr \; \ln (A)$, 
as:
\begin{eqnarray} \label{Seff-det}  
S_{eff}   &=&  - i \;  Tr  \; \ln \; \left\{ - 
i S_{q}^{-1} (x-y)
 \right\} 
,
\\
\label{Sc}
S_{q}^{-1} (x-y) &\equiv&
 S_{0}^{-1} (x-y) 
+ \Xi_s  (x-y)
+
\sum_q  a_q \Gamma_q j_q (x,y)  ,
\end{eqnarray}
where 
$Tr$ stands for traces of all discrete internal indices 
and integration of  spacetime coordinates
and $\Xi_s (x-y)$ stands for the coupling of  sea quark to 
the scalar-pseudoscalar fields
  for a particular 
pion field.
This  coupling term  can be written respectively for the
Weinberg pion field ($ \Xi_s^W  (x-y)$)
 and for the usual pion field ($\Xi_s^U  (x-y)$)
 in terms of unitary functions $U,U^\dagger$
as  \cite{EPJA-2016,EPJA-2018}:
\begin{eqnarray} \label{Xi-W}
 \Xi_s^W  (x-y) &=&
\left[ \gamma^\mu \vec{\sigma} \cdot {\cal D}_\mu \vec{\pi} i \gamma_5 
+ 
 i  \gamma^\mu \vec\sigma \cdot 
\frac{\vec{\pi} \times \partial_\mu \vec{\pi}}{1+{\vec{\pi}}^2} 
+  4  m  
\left( 
\frac{\vec\pi^2}{1+ \vec{\pi}^2}
-  \frac{\epsilon_{ijk} \sigma_k \pi_i \pi_j }{1+ \vec{\pi}^2}
\right)
\right] \delta (x-y) ,
\\    \label{Xi-U}
 \Xi_s^U  (x-y) &=& F ( P_R  U + P_L  U^\dagger) \; \delta (x-y)
,
\end{eqnarray}
 where $F=f_\pi$ 
is the pion field normalization, 
 $P_{R/L} =  (1 \pm \gamma_5)/2$ are the chirality 
 right/left hand projectors.

The free quark kernel can be written as 
$S_{0}^{-1} (x-y) = \left(  i \slashed{\partial} -  m 
\right) \delta (x-y)$,
where $m$ is so far the current quark mass.
The  classical solution for the scalar field, 
 found from  its gap equation,
 is directly incorporated into an effective quark mass 
 $M^* = m - <s>$.
The redefined quark kernel  can be written as:
\begin{eqnarray}
 S_{0}^{-1} (x-y)  = \left(  i \slashed{\partial} -  M^*
\right) \delta (x-y).
\end{eqnarray}
In expression (\ref{Sc}) the following quantity, with the usual chiral 
constituent quark currents that yield the leading couplings to pions,
 has been used:
\begin{eqnarray} \label{Rq-jch}
\frac{\sum_q  a_q \Gamma_q  j_q (x,y)}{ \alpha g^2}
&=&
2   R (x-y)
 \left[  \bpsi (y) \psi(x)
+ i  \gamma_5 \sigma_i  \bpsi (y) i \gamma_5  \sigma_i \psi (x)
\right]
\nonumber
\\
&-& 
 \bar{R}^{\mu\nu} (x-y) \gamma_\mu  \sigma_i \left[
 \bpsi (y) \gamma_\nu  \sigma_i \psi(x)
+  \gamma_5   \bpsi  (y)
 \gamma_5 \gamma_\nu  \sigma_i \psi (x) \right].
\end{eqnarray}
In this expression,
  $\alpha=2/9$ from the Fierz transformation,
$R(x-y)$ was given in (\ref{gluonpro-R}) and 
\begin{eqnarray} \label{gluonpro-Rbar}
\bar{R}^{\mu\nu}(x-y)  = g^{\mu\nu} (R_T (x-y) + R_L(x-y)) +
2 \frac{\partial^\mu \partial^\nu}{\partial^2} (R_T (x-y) - R_L(x-y) ).
\end{eqnarray}

\section{ Leading form Factors}

In the  following, consider the quark (and gluon) large effective mass expansion
for the   case in which
 quark and pion fields exchange momenta.
To provide the reader with one example, 
one of the leading pion constituent quark effective interactions
is the pseudoscalar coupling and it 
shows up  in the first order terms of the expansion as it follows:
\begin{eqnarray}
I_{det}^{ps} =  \frac{i}{2} \; Tr \;
\left[  S_0 (y-x)  i \gamma_5 \sigma_i   i \gamma_5 
\sigma_i
\pi_i (x)
S_0 (x-z) {R} (y-z) i\gamma_5  \sigma_j 
\bpsi(z)  i \gamma_5 \sigma_j \psi(y) \right] ,
\end{eqnarray}
With the insertion of complete sets of orthogonal momentum
states, 
a pseudoscalar
form factor at the constituent quark level emerges in 
momentum space,
$G_{ps}^U  (K,Q)$, where the  momenta $K,Q$
are defined below.
For this, the trace in internal indices (isospin, color and Dirac) were 
calculated.
By considering incoming quark  with momentum $K$,
and pion(s)   with total momenta $Q$  
the  set of  leading momentum dependent effective
 couplings  for the first pion definition (W) in the weak pion field limit 
($1 +  \vec{\pi}^2 \simeq 1$)
is given by:
\begin{eqnarray}  \label{L-Q-pi-W}  
{\cal L}^{q-\pi}_W &=&
M_3 (K) \; \bpsi (K)  \psi (K) 
\; + \; 2  
i \epsilon_{ijk} 
G_V^W (K,Q)
{\pi}_i  (q_{a})    \partial^{\nu}  {\pi}_j  (q_{b} )
\;  \bpsi (K) \gamma_\nu \sigma^k \psi (K+Q) 
\nonumber
\\
&+&  2 \;  
G_A^W (K,Q)  \;
  {\partial}^\nu {\pi}^i (Q)  \;  \bpsi (K) i \gamma_5 \gamma_\nu \sigma^i \psi (K+Q)
\; + \; 
F
G_{\beta sbF} (K,Q)
\; 
{\pi}_i (q_{1a}) \pi_i   (q_{b})
\;    \bpsi (K)   \psi (K+Q)
\nonumber
\\
&-& 
G_{ps}^{p,W} (K,Q) 
 \frac{\partial_\mu {\partial}^\mu {\pi}^i (Q)}{M^*}  
\;  \bpsi (K) i \gamma_5  \sigma^i \psi (K+Q)
- G_{s}^{p,W} (K,Q) 
\frac{ \partial_\mu {\partial}^\mu {\pi}^2 (Q)}{M^*}  \;
  \bpsi (K)  \psi (K+Q)
,
\end{eqnarray}
 where 
$Q=Q_\pi$ is the total momentum  carried by one or two  pion  in each of the vertices,
and it will be for both pion field definitions $W$ and $U$,
being that,  in the vector and scalar constituent quark currents couplings, 
$Q=q_{a}+q_{b}$
and the pion field was kept dimensionless.
The last two terms, momentum dependent ones, were obtained  with 
an integration by parts.
 In this expression $M_3(K)$ is a running effective mass 
that will be defined below in (\ref{M3}),
 and
the following dimensionless form factors  were defined
in terms of the functions $F_1(K,Q)$ given below:
\begin{eqnarray}
\label{GA-W}
G_A^W (K,Q) = G_V^W (K,Q)   &=&   
4 d_1 N_c (\alpha g^2)  \; F_1 (K,Q)
\\
G_{\beta sbF} (K,Q) &=& 
 64 d_1 N_c \frac{m}{F} (\alpha g^2) 
 \;
 F_1 (K,Q),
\\  \label{Gps-p-W} 
G_{ps}^{p,W} (K,Q)  = \frac{M^*}{4 m}  G_{s}^{p,W} (K,Q) 
&=&  16 {d_1}  {M^*} 
N_c (\alpha g^2)  \; F_2 (K,Q)
\end{eqnarray}
 where $N_c=3$,   $d_n= (-1)^{n+1}/(2 n)$.
It is interesting to note that the scalar pion coupling is proportional to the current quark 
mass and therefore it is a consequence of  explicit chiral symmetry breaking.
There are a scalar and a pseudoscalar momentum dependent form factors.
Although the usual pseudoscalar pion coupling to pseudoscalar quark current 
does not emerge at this level of calculation for the W pion field definition,
there is the coupling   $G^{p,W}_{ps}(K,Q)$ 
that might  contribute for the axial channel.
 Because it is simply proportional to other form factors by 
means of the function $F_2 (K,Q)$ it will not be investigated explicitely 
numerically below.
An analogous conclusion can be drawn for the derivative-scalar term
$G^{p,W}_{s}(K,Q)$ 
that might  contribute for the  vector channel.

The complete set of  leading momentum dependent couplings 
with their form factors  for the second   pion definition,
 with the same convention for momenta
of expression (\ref{L-Q-pi-W}) and dimensionless pion filed,
is given by:
\begin{eqnarray} \label{L-Q-pi-U} 
{\cal L}^{q-\pi}_U &=&   M_3 (K) \; \bpsi (K)  \psi (K) 
\; + \;     
G_{2js} (K,Q)
\; F
 {\pi}_i (q_{a}) \pi_i (q_{b})
   \bpsi (K)   \psi (K+Q)
\nonumber
\\
&+&   G_{ps}^U (K,Q)   F \;  {\pi}_i (Q)  
\;  \bpsi (K)  \sigma_i i \gamma_5 \psi (K+Q)
\nonumber
\\
&+&
  i \epsilon_{ijk}  \; 2 \;    G_V^U  (K,Q) \;
 \pi_i (q_{a})
 (\partial_\mu \pi_j (q_{b}) ) 
\;
 \bpsi  (K)   \gamma_\mu \sigma^j \psi (K+Q),
\nonumber
\\
&+&
  2 \;   G_{A}^U (K,Q) \;    (\partial^\mu \pi_i (Q)) 
\;
 \bpsi (K)  i \gamma_5 \gamma_\nu \sigma^i \psi (K+Q)
\nonumber
\\
&-& 
G_{ps}^p (K,Q) \frac{(\partial^2 \pi_i (Q))}{M^*}
\bpsi (K)  \sigma_i i \gamma_5 \psi (K+Q) 
-
G_{s}^p (K,Q) \frac{\partial^2 (\pi_i (q_a) \pi_i (q_b) )}{M^*}
\bpsi (K)   \psi (K+Q)
,
\end{eqnarray}
where $M_3(K)$ is the same as the mass 
in expression (\ref{L-Q-pi-W})
 and it will be  defined in expression (\ref{M3}).
The other form factors were defined as:
\begin{eqnarray}  \label{Gps}
G_{ps}^U (K,Q) &=&  G_{2js} (K,Q)   = 
 32 d_1 N_c  (\alpha g^2)  \;   F_1(K,Q)  ,
\\  \label{GA-U} 
 G_A^U (K,Q) &=&  G_V^U (K,Q)    = 
16 d_1 N_c F \; (\alpha g^2)  \;    F_{2}(K,Q) 
\\
\label{gps-k}
G_{ps}^p (K,Q) &=& G_{s}^p (K,Q) = 16 d_1 N_c \;
F \; (\alpha g^2) 
F_2(K,Q)
,
\end{eqnarray}
The derivative couplings with 
form factors 
 $G^p_{ps}(K,Q)$  and  $G^p_s(K,Q)$ 
 have simply a different normalization with respect to 
 the ones from the  W pion field definition:
$G^{p,W}_{ps}(K,Q)$ and $G^{p,W}_s(K,Q)$.
For example,  it can be seen that 
$G^{p,W}_{ps}(K,Q) = \frac{M^*}{F} G^{p}_{ps}(K,Q)$.
At this level, it is interesting to note that 
$G_{ps} (K,Q) =  G_{2js} (K,Q)$
in reasonable agreement
with other results \cite{craig-bloch-schimidt},
and also $G_A(K,Q)=G_V(K,Q)$ for both 
pion field definitions.

The loop momentum  integrals of each of the form factors above 
 will  be 
written and investigated    for  constituent quark with $K=0$, except for
the effective mass $M_3(Q)$. 
After a Wick rotation 
for the Euclidean momentum space
these functions are given by:
\begin{eqnarray} \label{F1-Q}
F_1(0,Q) &=& 
\int_k ( k \cdot (k+Q) - {M^*}^2 ) \tilde{S}_0(k) \tilde{S}_0 (k+Q) \bar{\bar{R}}(-k)
,
\nonumber
\\
F_2 (0,Q)
 &=&
\int_k   {M^*}
 \tilde{S}_0(k) \tilde{S}_0 (k+Q) \bar{\bar{R}}(-k)
\nonumber
\\
\label{M3}
M_3 (Q)
 &=&
16 d_1 N_c M^*  (\alpha g^2)
 \int_k \tilde{S}_0 (k+Q) R (-k) 
,
\end{eqnarray}
where $\int_k = \int \frac{d^4 k}{(2 \pi)^4}$
and the following functions  
in momentum space for components of the 
quark and gluon propagator used:
\begin{eqnarray}
\tilde{S}_0 (k) &=& \frac{1}{ k^2 + {M^*}^2},
\\
\bar{\bar{R}} (k) &=& 2 R (k) =
 6 R_T (k) + 2  R_L (k),
.
\end{eqnarray}
The only form factor that might have an ultraviolet divergence UV is
$M_3(Q)$ if the gluon propagator does not possess
particular UV behavior.
The other   are completely 
finite if the non perturbative gluon propagator is infrared regular.

The momentum structure of the form factor
$F_1(0,Q)$  has a positive first derivative with respet to $Q^2$
for very small $Q$,
and therefore it yields negative quadratic radii. 
To overcome that, $F_1(0,Q)$
might be truncated by 
 approximating the quark kernel
 by
$S_0(k) \simeq M^* \tilde{S}_0(k)$.
It yields for the function $F_1(0,Q)$ the following expression:
\begin{eqnarray} \label{F1-tr}
F_1^{tr}(0,Q) &=& {M^*}^2 
\int_k \tilde{S}_0 (k) \tilde{S}_0 (k+Q) \bar{\bar{R}}(-k) 
.
\end{eqnarray}
 This truncation might be expected to 
 correspond to making an effective mass $M^*$ to 
be momentum dependent in the  expression of $F_1(K,Q)$.

\begin{figure}[ht!]
\centering
\includegraphics[width=120mm]{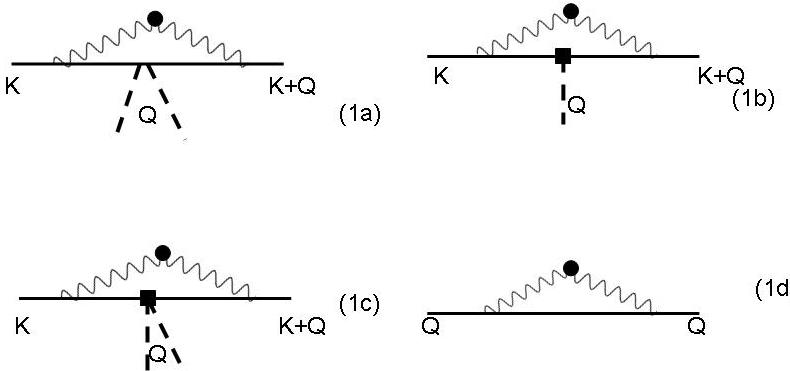}
\caption{ \label{diagrams-W-1}
\small
These diagrams   correspond  to the quark-pion effective couplings
from expression (\ref{L-Q-pi-W}).
 The wavy line with a full dot is a (dressed) non perturbative gluon propagator,
the solid lines stand for a constituent quark (external line) or sea quark (internal line), 
and   dashed lines represents   pion field,
the full square in a vertex represents a derivative coupling.
Diagram  (1d) represents  the effective quark mass correction.
}
\end{figure}

\begin{figure}[ht!]
\centering
\includegraphics[width=100mm]{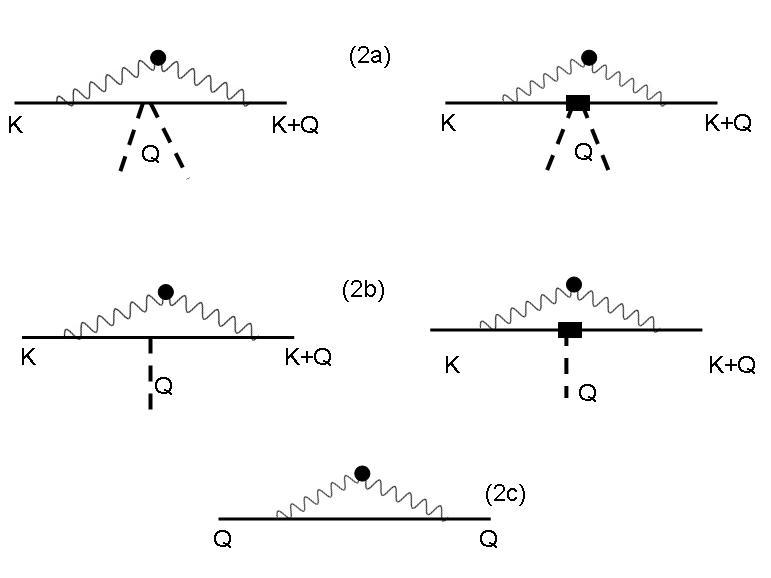}
\caption{ \label{diagrams-2}
\small
In these diagrams, the wavy lineThese diagrams   correspond  to the quark-pion effective couplings
from expression (\ref{L-Q-pi-U}).
 The wavy line with a full dot is a (dressed) non perturbative gluon propagator,
the solid lines stand for a constituent quark (external line) or sea quark (internal line), 
and   dashed lines represents   pion field,
the full square in a vertex represents a derivative coupling.
}
\end{figure}

 In Figure (\ref{diagrams-W-1}), the diagrams corresponding to the expressions 
 (\ref{L-Q-pi-W}) for the Weinberg pion field definition
 are presented, where the pion-quark vertices with a 
square are the derivative ones and diagram (1d)
stands for the effective mass $M_3(Q)$.
 The dressed (non perturbative) gluon propagator is indicated
by a wavy line with a full circle
and pion is represented by dashed lines. 
In diagrams (1a-c) he incoming constituent has momentum $K$ and the outgoing constituent quark 
has momentum $K+Q$, being $Q$ the total momentum transfered by 
pion(s).
 Figure (\ref{diagrams-2}) exhibits the diagrams for the 
pion constituent quark couplings for the  usual pion field definition 
given in expression (\ref{L-Q-pi-U})
 with the same conventions  of Figure 1.

\section{ Numerical results}

To provide numerical results, 
two gluon propagators were chosen.
A transversal  one  from Tandy-Maris $D_{I}(k)$
\cite{tandy-maris}
and  the other is an  effective longitudinal confining one  by Cornwall 
$D_{II}(k)$
 \cite{cornwall}.
Both of them 
yield DChSB
and they  are written below with 
the following association:
\begin{eqnarray} \label{propagators}
g^2 \tilde{R}^{\mu\nu} (k) \equiv  h_a D_a^{\mu\nu} (k)
\end{eqnarray}
where $D^{\mu\nu}_a(k)$ ($a=I,II$) 
 is one of the chosen  gluon propagators
from the  quoted articles, $h_a$ is a real positive 
 constant factor
used in previous works \cite{EPJA-2018,PRD-2018b}
to fix the quark gluon (running) 
coupling constant such as
to reproduce one  expected value 
either of the  vector/axial 
 pion coupling  constant in the vacuum or vector meson coupling to constituent quarks
constant,
$g_V h_a = 1$, $g_A h_a = 1$  or $g_\rho h_a \simeq 12$.
In the present work 
this factor was chosen for each of the gluon propagators
and pion field definition to provide $g_A (0) h_a = 1$. 
 Their values will be shown in the caption of 
 the corresponding figure.

The expressions for the two  gluon propagators 
 are the following:
\begin{eqnarray}
D_I (k) &=& 
\frac{8  \pi^2}{\omega^4} De^{-k^2/\omega^2}
+ \frac{8 \pi^2 \gamma_m E(k^2)}{ \ln
 \left[ \tau + ( 1 + k^2/\Lambda^2_{QCD} )^2 
\right]}
,
\\
D_{II} (k) &=& 
\frac{K_F}{(k^2+ M_k^2)^2 },
\end{eqnarray} 
where
for the first expression 
$\gamma_m=12/(33-2N_f)$, $N_f=4$, $\Lambda_{QCD}=0.234$GeV,
$\tau=e^2-1$, $E(k^2)=[ 1- exp(-k^2/[4m_t^2])/k^2$, $m_t=0.5 GeV$,
$\omega = 0.5$GeV, $D= 0.55^3/\omega$ (GeV$^2$); 
and for the second expression
$K_F = (  2 \pi  M_k / (3 k_e) )^2$
where  $k_e  = 0.15$   and $M_k = 220$MeV.

\begin{figure}[ht!]
\centering
\includegraphics[width=140mm]{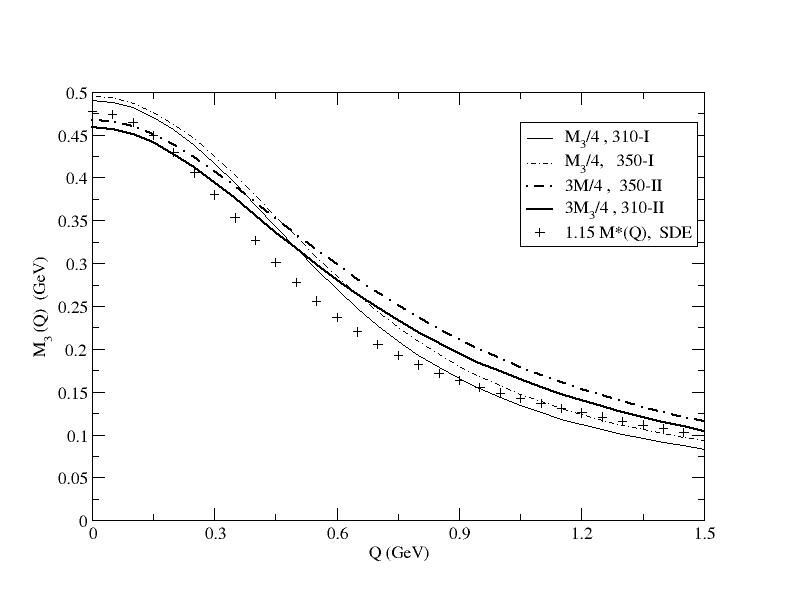}
\caption{ \label{fig:M3}
\small
The dynamical  running  constituent quark effective mass $M_3(Q)$
divided by $4$, $M_3(Q)/4$, for the gluon propagator $I$, in continuous and 
dashed thin lines and multiplied 
by $3/4$, $\frac{3}{4}M_3(Q)$, in thick lines  for the gluon propagator $II$,
in dashed ($M^*=350$MeV) and continuous ($M^*=310$MeV) lines.
A gap effective mass $M^*(Q)$ 
 from SDE from Ref. (\cite{craig-mass}) multiplied by a factor
1.15 to allow for a better comparision of the momentum dependence.
}
\end{figure}

In Figure (\ref{fig:M3}) the resulting constituent quark (running) effective 
mass $M^*_3(Q)$ is shown as a function of the constituent quark momentum
for an UV cutoff $\Lambda=2$GeV, 
in dashed and continuous  
lines and it is compared to a result from Schwinger Dyson equations
at the rainbown ladder approximation from Ref. \cite{craig-mass}.
The multiplicative factors $1/4$ and $3/4$ were chosen to fit the 
curves into a suitable scale and they are needed because of 
the large value of $\Lambda$.

\begin{figure}[ht!]
\centering
\includegraphics[width=140mm]{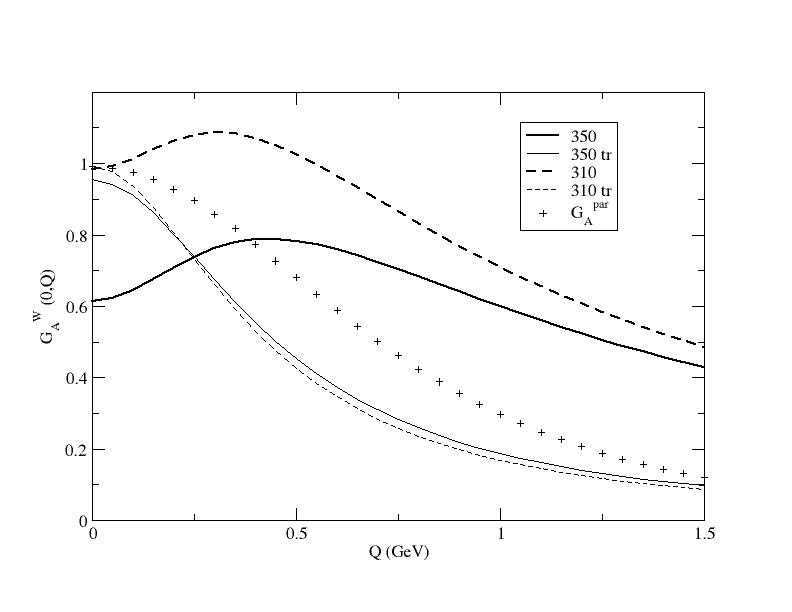}
\caption{ \label{fig:GAW-CO}
\small
The normalized  axial 
 form factor $G_A^W(0,Q)$  
for the Weinberg pion field
  as a function of the pion momentum 
 is presented in this figure
for 
the gluon propagator $D_{II}(k)$, untruncated  expression with 
$h_a =\frac{1}{0.2}$ and truncated one with  $h_a=\frac{1}{0.46}$,
and  for different values of the
sea quark effective mass $M^*$ from the gap equation.
 Solid thin and thick   lines  for $M^*=350$MeV;
dashed lines for $310$MeV.
Signs $+$ for the normalized  fitting of expression (\ref{gA-par}), $G_A^{par}(Q^2)$. }
\end{figure}

\begin{figure}[ht!]
\centering
\includegraphics[width=140mm]{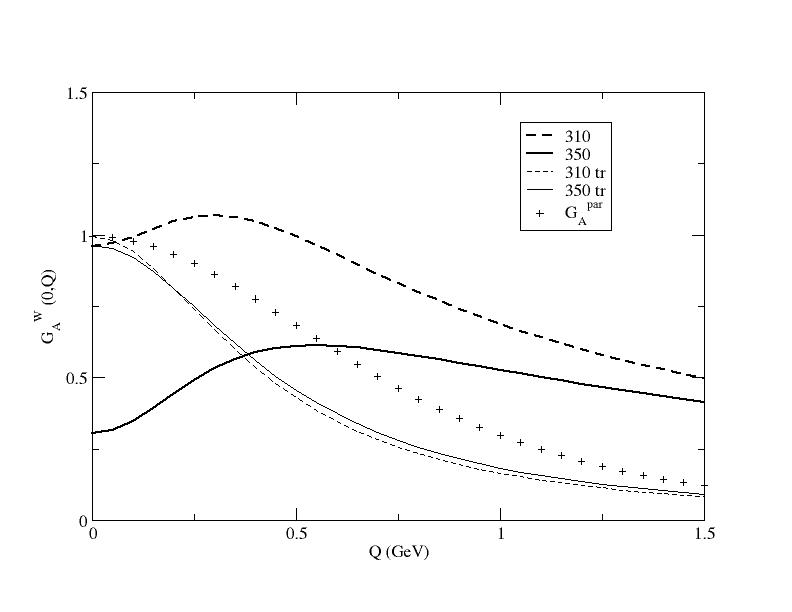}
\caption{ \label{fig:GAW-TM}
\small
The axial 
 form factor $G_A^W(0,Q)$  
for the Weinberg pion field
  as a function of the pion momentum 
 is presented in this figure
for 
the gluon propagator  $D_{I}(k)$ from reference \cite{tandy-maris},
being that  for the   untruncated expression it is taken
$h_a =\frac{1}{0.38}$ and for the  truncated 
one $h_a=\frac{1}{1.4}$,
and  for different values of the
sea quark effective mass $M^*$ from the gap equation.
 Solid thin and thick   lines  for $M^*=350$MeV;
dashed lines for $310$MeV.
Signs $+$ for the fitting of expression (\ref{gA-par}), $G_A^{par}(Q^2)$.
 }
\end{figure}

In Figure (\ref{fig:GAW-CO})
the axial form factor contribution 
 for  zero quark momentum $G_A^W(0,Q)$
and its truncated version  $G_A^{W,tr} (0,Q)$
for the Weinberg pion field are  presented for different values of 
the quark effective mass from the gap equation $M^*$ and 
for the  gluon propagator $D_{II}(k)$.
 In all cases of the figures with the axial form factor, the 
linear dependence on the pion momentum from the coupling
 was  not included.
In Figure (\ref{fig:GAW-TM}) the same results are exhibitted for 
the gluon propagator $D_I(k)$.
Figures (\ref{fig:GAW-CO}) and (\ref{fig:GAW-TM})
present the same behavior without meaninful differences except for the 
relative normalization
of the non truncated form factor.
Besides that,
a  dipolar  fitting for 
 experimental  results of axial pion-nucleon coupling 
 is drawn with symbols $+$ with a normalization to 
allow for comparison of the momentum dependence.
It is given by 
\cite{exp-bernard+E+meissner,PDG,alexandrou-etal}: 
\begin{eqnarray} \label{gA-par}
G_A^{par} (Q^2) &=& \frac{G_{0}}{ \left( 1 + \frac{Q^2}{M_A^2} \right)^2 }, 
\end{eqnarray}
by considering $M_A=1.1$ GeV 
and by adopting a normalization for $G_A^{par} (Q^2=0)$ obtained 
in the present work for each of the gluon propagators, for the case of 
$M^*=0.31$GeV.
The fitting for experimental values decreases  slower than
the (constituent quark) form factors $G_A^W(0,Q)$ 
and two reasons might directly  identified for that.
 It  might  signal 
there is missing strength from  more complete 
quark and gluon kernels.
However it  also might indicate   the need to account other
effects rather related to 
 nucleon structure
 degrees of freedom.
These two possibilities are not excludent, however they
correspond to different types of constituent quark models for 
hadrons (baryons) since they would correspond to 
different roles of constituent quark interactions for 
the baryon structure.
In any case, apart from 
a possible difference on the overall normalization,
  the difference is not very large and it appears in intermediary momenta.
It can be noted that the non truncated expressions provide
a positive momentum slope at $Q=0$, these expressions therefore
would provide a negative averaged quadratic  axial  radii.
The truncated expressions correct this behavior.

\begin{figure}[ht!]
\centering
\includegraphics[width=140mm]{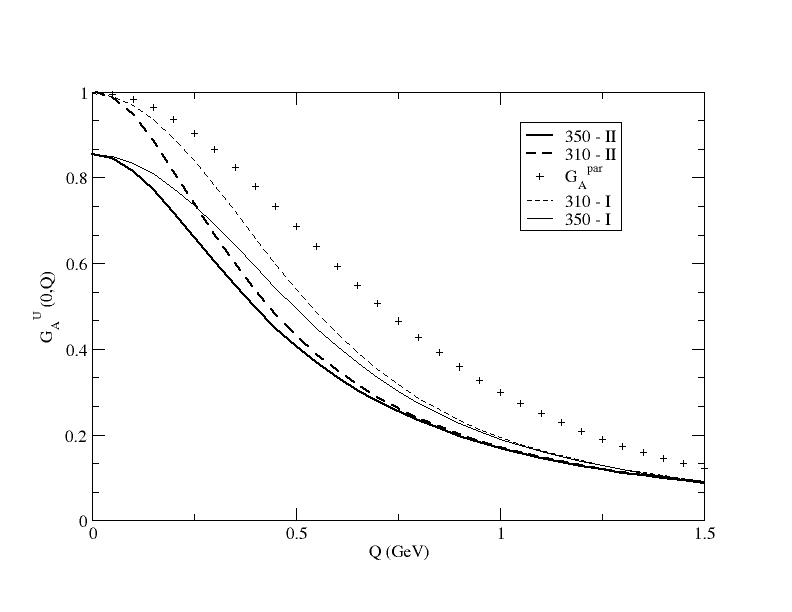}  
\caption{ \label{fig:gA-U-CO}
\small
The axial (equal to the vector)
 form factor $G_A^U(0,Q)$ (usual pion field) 
  as a function of the pion momentum 
 is presented in this figure
for 
the gluon propagators $D_I(k)$,
with factor
$h_a =\frac{1}{0.83}$,
and 
 $D_{II}(k)$, with 
$h_a =\frac{1}{0.27}$, 
and  for different values of the
sea quark effective mass $M^*$, from the gap equation.
Solid line is  used for $M^*=350$MeV, dashed line for $M^*=310$MeV.
Sign $+$ for the fitting of expression (\ref{gA-par}), $G_A^{par}(Q^2)$.
}
\end{figure}

In Figure (\ref{fig:gA-U-CO})
the axial form factor  correction for the second pion field definition,
$G_A^U(0,Q)$, 
as a function of pion momentum
is shown  for gluon propagators
$D_{II}(k)$ and $D_I(k)$  for different values of 
the quark effective mass $M^*$.
The same fitting $G_A^{par}(Q)$ is plotted  ($+$)
with the value at $Q=0$ adjusted from the 
$G_A^U(0,Q=0)$ to make an appropriated  
comparison.
The truncated version of $G_A^{W,tr}(0,Q)$ 
from figures (\ref{fig:GAW-CO}) and (\ref{fig:GAW-TM})
have  a similar  behavior to
$G_A^U(0,Q)$, and in fact they both are written in terms of 
$F_2(0,Q)$
with different normalizations
Although the overal behavior is similar 
to the experimental fit,
for both $G_A^{W,tr}(0,Q)$ and $G_A^U(0,Q)$,
the form factor contribution  $G_A^U(0,Q)$ has a behavior 
slightly closer to the experimental fit.

The axial coupling constant  
at the constituent quark level has been
argued to be close to $g_A \simeq 3/4$ 
\cite{weise-vogl}
or $g_A \simeq 1$ \cite{weinberg-2010}.
Results from the form factors are very well of the 
correct order of magnitude and value.
 Also, 
in the present work, it was shown in expressions
(\ref{GA-W})  and (\ref{GA-U}) 
the axial and vector form factors are equal to each other,
due to chiral symmetry,
 for each the two pion field
definitions considered.

\subsection{Pseudoscalar coupling}

\begin{figure}[ht!]
\centering
\includegraphics[width=140mm]{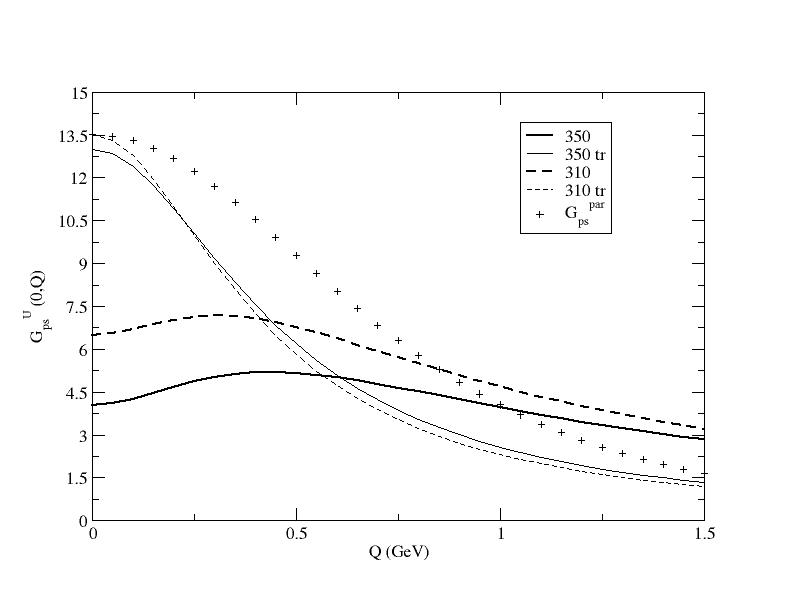}  
\caption{ \label{fig:gps-u-co}
\small
The pseudoscalar form factor $G^U_{ps}(0,Q)$ 
as a function of the quark momentum 
 is presented in this figure
for 
the gluon propagator $D_{II}(k)$, with factor
$h_a =\frac{1}{0.27}$,
and  for different values of the
sea quark effective mass $M^*$, from the gap equation.
Results from both the complete and the truncated (tr) expressions 
are shown.
Solid lines are used for $M^*=350$MeV, dashed lines for $M^*=310$MeV.
Signs $+$ for the corresponding fitting of expression (\ref{gA-par}), $G_A^{par}(Q^2)$. 
}
\end{figure}

\begin{figure}[ht!]
\centering
\includegraphics[width=140mm]{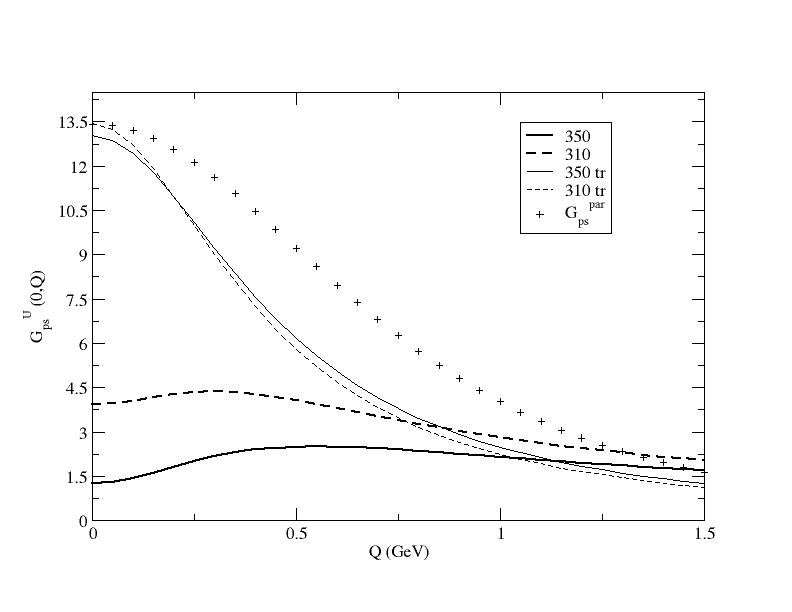} 
\caption{ \label{fig:gps-u-tm}
\small
The pseudoscalar form factor $G^U_{ps}(0,Q)$ 
as a function of the quark momentum 
 is presented in this figure
for 
the gluon propagator $D_{I}(k)$, with factor
$h_a =\frac{1}{0.83}$,
and  for different values of the
sea quark effective mass $M^*$, from the gap equation.
Results from both the complete and the truncated (tr) expressions 
are shown.
Solid lines are used for $M^*=350$MeV, dashed lines for $M^*=310$MeV.
Signs $+$ for the corresponding fitting of expression (\ref{gA-par}), $G_A^{par}(Q^2)$. 
}
\end{figure}

In figures (\ref{fig:gps-u-co}) and (\ref{fig:gps-u-tm})
the pseudoscalar form factor $G_{ps}^U(0,Q)$
and its truncated version  $G_{ps}^{U,tr}(0,Q)$
are  presented  for the 
gluon propagators $D_{II}(k)$ and 
$D_{I}(k)$ respectively.
The zero momentum $Q=0$ values 
 are  basically  one order of magnitude  
larger than   the zero momentum 
axial form factor as expected from phenomenology.
Results with $D_I(k)$ have  considerably larger absolute values than 
 wtih $D_{II}(k)$.
The dipolar fitting for data from lattice QCD calculations  (\ref{gA-par})
 \cite{sasaki-yamazaki} 
is  also shown  with a suitable  normalization at $G_{ps} (0,0)$ 
 to compare with the results from expressions above 
for the case $M^*=0.31$GeV.
All the results from the truncated expressions for 
$G_{ps}^U(0,Q)$  yield similar results 
for $M^*= 0.31$ and  $0.35$GeV.
Whereas the truncated version presents 
 a monotonic decrease with 
momentum $Q$ the complete expression has an increase up to around 
$Q\sim 0.40-0.45$GeV and 
then it decreases
for larger $Q$.
 It has therefore the same behavior of 
$G_A^W(Q)$ shown in the previous section.
The deviation of the form factor  $G_{ps}^{U,tr}(0,Q)$
 momentum dependence from the 
fitting $G_{ps}^{par}(0,Q)$
 is slightly larger than 
the deviation of the axial $G_A^{U} (0,Q)$ form factor
with respect to the corresponding nucleon-pion
 experimental fitting.
The reasons must be the same, the momentum dependence of the 
quark and gluon kernels and/or internal nucleon effects.

Standard hadron effective coupling constants 
are usually  obtained for  particular values  of the 
transfered momentum  such as $Q^2=0$ or $Q^2\simeq-m_\pi^2$.
The only numerical values for the form
factors at the  timelike momenta $Q^2< 0$  shown in this work
 are these next ones for the  usual pseudoscalar pion coupling 
at $Q^2= - m_\pi^2$,
 i.e. closer to the physical definition of $G_{\pi N}$
that is taken from  timelike momenta at the 
muon or pion mass.
For the quark  effective mass $ M^*=0.31$GeV
 and the two gluon propagators
 two values were obtained:
for the complete  expression (\ref{Gps})
 and for the momentum truncated expression
$G_{ps}^{W,tr}(0,Q)$ with 
(\ref{F1-tr}).
By considering the same factors $h_a$
adopted for the figures  of the pseudoscalar form factors
($h_I=1/0.83$ and $h_{II}=1/0.27$),
 they are given by:
\begin{eqnarray}
 I \; 
\;\;\;\;\;\; 
G_{ps}^U (0,Q^2=-m_\pi^2) &=& 1.9,
, \;\;\;\;\; G_{ps}^U (0,0) =  3.4,
\\
 I \;
\;\;\;\;\;\;
G_{ps}^{U,tr} (0,Q^2=-m_\pi^2) &=& 16.4,
, \;\;\;\; G_{ps}^{U,tr} (0,0) =  13.4, 
\\
 II \;  
\;\;\;\;\;\;
G_{ps}^U (0,Q^2=-m_\pi^2) &=&   4.1;
, \;\;\;\;\; G_{ps}^U (0,0) =  5.9,
\\
II \;  
\;\;\;\;\;\;
G_{ps}^{U,tr} (0,Q^2=-m_\pi^2) &=& 15.2,
, \;\;\;\;\; G^{tr}_{ps} (0,0) =  13.3.
\end{eqnarray}
The difference between the form factor $G_{ps}^U(0,Q)$
and its truncated version, $G_{ps}^{U,tr}(0,Q)$, is of course present in 
these  timelike  values.
The values from the truncated expression are also closer to experimental 
data  for the nucleon-pion coupling constant and 
results from  other calculations.

\subsection{Goldberger Treiman and other relations in  spacelike  momenta}

Next 
 ratios of the form factors are calculated.
The  following  momentum dependent ratios 
 between dimensionless quantities
were considered:
\begin{eqnarray}
GT_W(Q) 
&\equiv& \frac{G_V^W (0,Q) }{ G_{\beta sbF} (0,Q) }  
=
\frac{F}{16 m} 
 \simeq 1 ,
\\
GT (Q) &=& \frac{M^*}{F} \left(
\frac{ G_{A}^U (0,Q)   }{ G_{ps}^U (0,Q)  } \right)
=
\frac{M^*}{F}
\frac{G_{V} (0,Q)}{ G_{2js} (0,Q)} = 
\frac{M^*}{F}
\frac{F}{2}    \frac{  F_2 (0,Q) }{ F_1(0,Q) }
,
\\
\frac{ G_{A}^U (0,Q)  }{G_{V}^U (0,Q)  } &=&
 \frac{ G_{ps}^U (0,Q)   }{ G_{2js} (0,Q)   }
= 1 ,
\end{eqnarray}
where the first one ${GT_W}(Q)$ is an equivalent of the 
GTR expression for the Weinberg pion field in which the pseudoscalar 
pion coupling does not appear but the (symmetry breaking) 
scalar two pion coupling to constituent quark appears.
This  ratio  is momentum independent and it depends on the 
current quark mass $m \sim 5.75$MeV for which $16 m \simeq f_{\pi} = F= 92$MeV and 
therefore  ${GT_W}\simeq 1$.
The function $GT(Q)$  for the second pion definition
has a  constant factor  $F/M^*$
such that if the GTR relation is satisfied the ratio $GT(Q)\to 1$
and  this is  verified for very  large $M^*$.
The last expression  has two  chiral symmetry 
  relations for  form factors, and their corresponding 
effective coupling constants
   for the second pion field definition.

\begin{figure}[ht!]
\centering
\includegraphics[width=140mm]{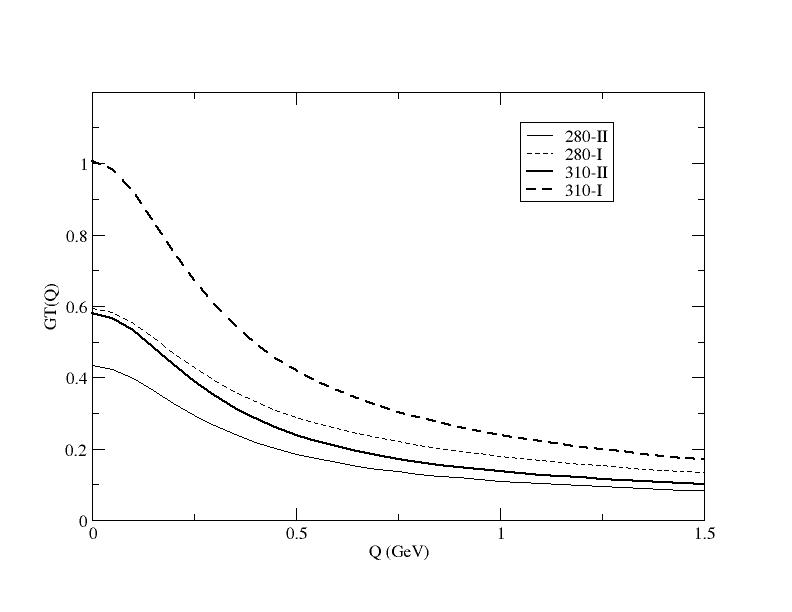}
\caption{ \label{fig:GT-Q}
\small
The ratio $GT(0,Q)$ is shown as a function of momenta for two different
quark effective masses $M^*=280, 310$MeV and for the two gluon propagators.
The limit in which the 
Goldberger Treiman relation is recovered corresponds to 
$GT(0)=1$.
}
\end{figure}

\begin{figure}[ht!]
\centering
\includegraphics[width=140mm]{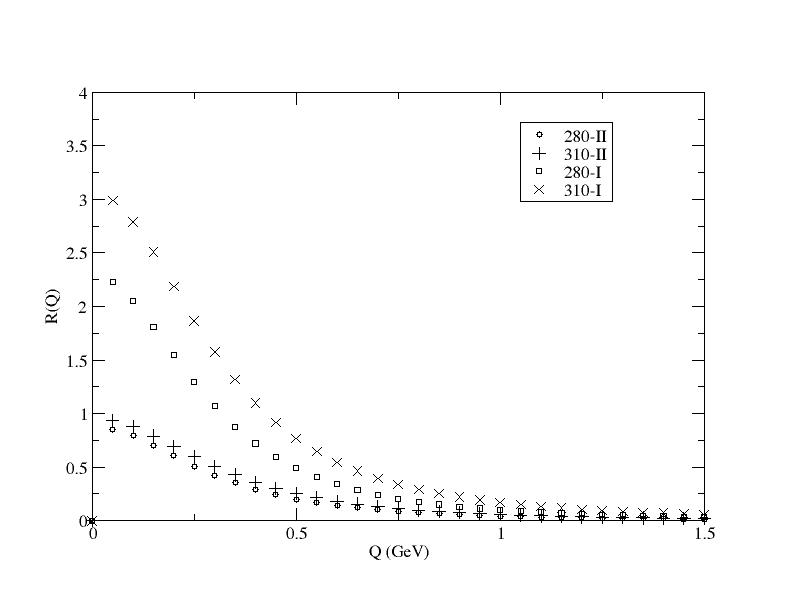}
\caption{ \label{fig:Rq-normaliz-new}
\small
The momentum dependent deviation of the 
Goldberger Treiman relation
for 
the gluon propagators
$D_I(k)$ and  $D_{II}(k)$
and quark effective masses $M^*=280$ and $310$MeV,
by considering normalized definitions for $G_A^U(0,0)$ and $G_{ps}^U(0,0)$
such as to  satisfy the GTR at $Q=0$.
}
\end{figure}

In figure  (\ref{fig:GT-Q}) 
  the ratio  $GT(Q)$ is presented
as a function of momentum
for different effective quark masses $M^*$. 
The ratio $GT(Q)$ 
 does not satisfy necessarily the GTR  at $Q=0$ because
the quark effective masses are not large enough.
This ratio $GT(Q)$ 
 has the same behavior found in  other works
 \cite{eichmann-fischer}.
 The deviation from the GTR  intrinsically due to the momentum dependence of
each of the form factors for the nucleon level  Goldberger-Treiman relation 
is usually  denoted by $R(Q)$.
It is usually 
parameterized in terms of the nucleon mass $M$  \cite{eichmann-fischer}, and by
substituing $M$ by the quark effective mass $M^*$
 it is given by the following expression:
\begin{eqnarray}
G_A(Q^2) &=& \frac{f_\pi}{M^*} 
G_{\pi NN}(Q^2) - \frac{Q^2}{4M^*} R(Q^2),
\end{eqnarray}
where $G_{\pi N}(Q^2)$ is to be substituted by $G_{ps}(Q)$.
By considering the constituent quark mass $M^*=0.28$GeV and $0.31$GeV
  this function is exhibitted in figure (\ref{fig:Rq-normaliz-new})
for   the second pion definition.
 It goes to zero quite fast  with increasing $(Q)$ depending not only 
on the quark effective mass $M^*$ but also on the gluon propagator considered.

\subsection{Averaged  quadratic  radii}

Next, the corresponding    strong
averaged
quadratic   radii are defined 
from the different pion-constituent quark  couplings
presented  above.
Since the form factors are dimensionless the corresponding 
axial and pseudoscalar  quadratic radii 
were defined   by:
\begin{eqnarray}   \label{rA-W}
< r^2 >_{A}^W &=&
-  6  \left. 
  \frac{ d G_A^W (0,Q)
 }{d Q^2 } \right|_{Q=0}  = < r^2 >_V^W
,
\\ \label{r-A-W-tr}
< r^2 >_{A}^{W,tr}&=&
-    6   \left.   
 \frac{ d G_A^{W,tr}  (0,Q)}{d Q^2 } \right|_{Q=0}
 = < r^2 >_V^{W,tr} 
,
\\ \label{r-A-U}
  < r^2 >_A^U  &=&
- 
 6 \left.       \frac{ d G_{A}^U (0,Q)}{d Q^2 } \right|_{Q=0}
= < r^2 >_V^U ,
\\   \label{r-ps-U}
 < r^2 >_{ps} &=&
-  
6  \left.   \frac{ d G_{ps}^U (0,Q)}{d Q^2} \right|_{Q=0}
 =  < r^2 >_{2js}
,
\\   \label{r-ps-U-tr}
 < r^2 >_{ps}^{tr} &=&
-  
6   \left.    \frac{ d G_{ps}^{U,tr} (0,Q)}{d Q^2 } \right|_{Q=0}
 =  < r^2 >_{2js}^{tr}
.
\end{eqnarray}
where in the right hand side of these expressions
 the relations  to vector and scalar 
 quadratic radii from the  
form factors defined in the previous sections
are exhibitted.
In \cite{PRD-2018a} the light vector/axial mesons couplings to constituent quarks
were considered to provide  corresponding quadratic radii.
The corresponding 
averaged
  axial and vector quadratic  radii 
seen by the coupling to the pion, presented in this work, also 
turn out to be equal.
Both results, from the pion and axial mesons couplings,
 are to be added, i.e.
in fact expressions (\ref{rA-W}-\ref{r-A-U}) 
provide corrections to the corresponding 
quadratic radii.
However their  experimental values, 
at the nucleon level, must receive further corrections since
vector and axial a.q.r. are different from each other
and  expected to  follow:
$\sqrt{ <r_V^2>/<r_A^2> } \simeq 1.6$ \cite{weise-vogl}.

\begin{figure}[ht!]
\centering
\includegraphics[width=140mm]{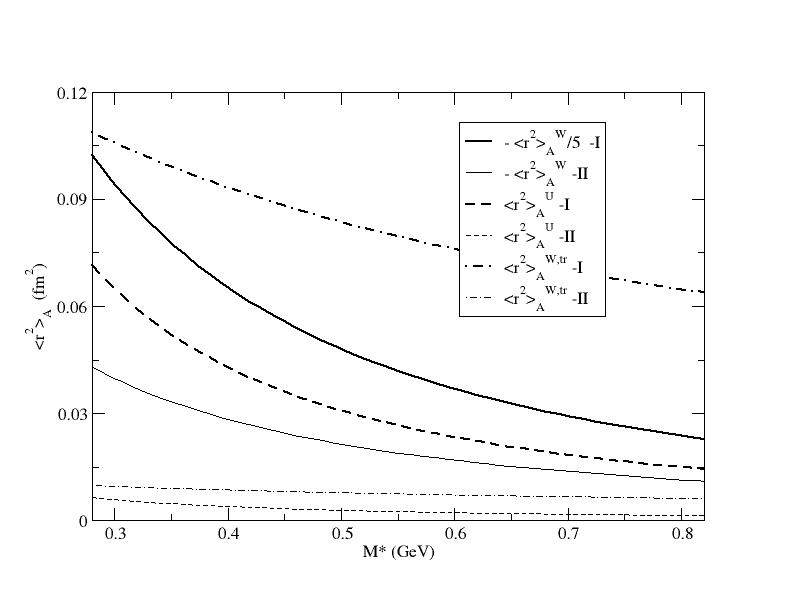}
\caption{ \label{fig:rqm-A}
\small
The axial quadratic averaged  radius (contribution)
 for the two pion definitions, $W$ and $U$,
and two gluon propagators, $I$ and $II$,
 as functions of the effective quark mass $M^*$.
The factors $h_a$ 
were chosen to be $h_I=3$ and $h_{II}=1$.
The numerical result for $<r^2>_A^W$
 has a sign minus and the results for the gluon propagator $D_I$
 it is divided by $5$ to 
fit in  the scale of the figure.
}
\end{figure}

In figure (\ref{fig:rqm-A}) the 
different estimations for the axial quadratic radius contribution for 
the two pion definitions, $W$ and $U$, and for the  two 
gluon propagators,
 as functions of the quark effective mass $M^*$.
 In the figures with a.q.r.
the  factors $h_a$ 
 were considered $h_I=3$  and $h_{II}=1$, 
such that results could be compared with results from \cite{PRD-2018a}.
In the case of the Weinberg definition
there are  also results for the truncated
expression.
The axial radius   (contribution) 
 $<r^2>_A^W$ is negative because of the behavior of the 
axial form factor close to zero exchanged
 momentum and this
 unexpected behavior is corrected by the truncated expression
as discussed above.
Besides the problem with the sign for $<r^2>_A^W$ 
it is also noted a  different behavior
in the $M^*$-dependence of the axial quadratic radii
between $<r^2>_A^W$  and $<r^2>_A^{W,tr}$ ,
being that the former presents a  stronger variation for increasing $M^*$
and the latter a smoother variation.

\begin{figure}[ht!]
\centering
\includegraphics[width=140mm]{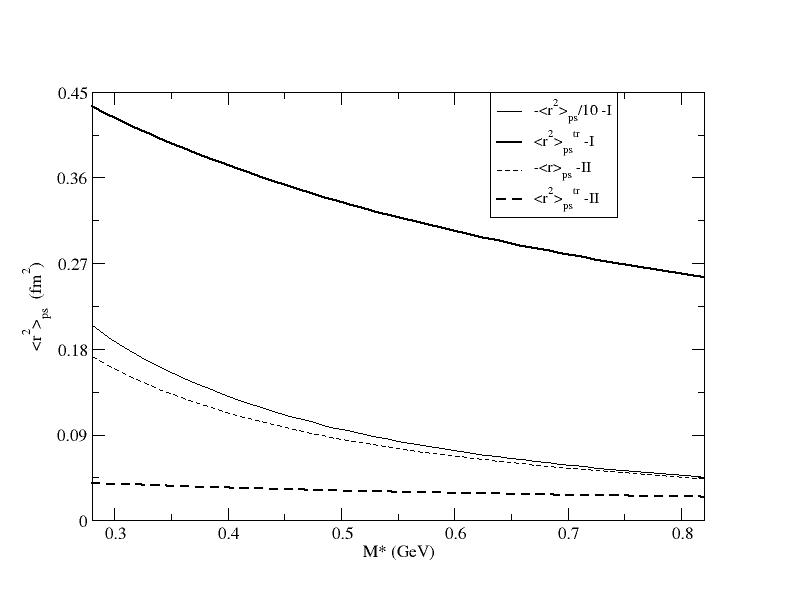}  
\caption{ \label{fig:rqm-ps}
\small
The pseudoscalar averaged   quadratic
  radius
and two gluon propagators, $D_I$ and $D_{II}$,
 as functions of the effective quark mass $M^*$, with factors
$h_a$ from the previous figures.
The numerical result for $<r^2>_{ps}$
 has a sign minus and the results for the gluon propagator $D_I$
 it is divided by $10$ to 
fit in  the scale of the figure.
The factors $h_a$ 
were chosen to be $h_I=3$ and $h_{II}=1$.
}
\end{figure}

These axial quadratic radii correction due to the 
pion  are smaller than the vector/axial quadratic radii due to 
the vector/axial light mesons calculated with the same method 
for both gluon propagators
 in \cite{PRD-2018a}.
In that work the axial quadratic radii 
found from the coupling to the $A_1$ meson,
$<r^2_{a.m.}>_A$,
 were estimated  to be in the 
following range of values - for the same range of values of the quark
effective mass $M^*$ - by keeping the corresponding $h_a$
to the ones used in the figures for the a.q.r.,:
\begin{eqnarray}  \label{rqm-vecmes}
<r^2_{a.m.}>_A &\sim& 0.4 - 0.2 \;  \mbox{fm}^2, \;\;\;\;\;\;\; D_{II}(k),
\nonumber
\\
<r^2_{a.m.}>_A &\sim& 4.0 - 2.0 \;  \mbox{fm}^2, \;\;\;\;\;\;\; D_I (k),
\end{eqnarray}
  respectively 
for gluon propagators $D_{II}(k)$ and $D_{I}(k)$.
Of course the estimations for $<r^2_{a.m.}>_A$ with $D_I(k)$ 
are 
extremely large, also present in figures (\ref{fig:rqm-A},\ref{fig:rqm-ps})
 and this had been attributed rather to the corresponding
quark-gluon coupling constant and gluon propagator strengths.
Both resulting values however 
 are basically of the order  of  magnitude 
as (or larger than)
 the estimation for  constituent quark radius
 $\sqrt{<r^2>_{CQ}} \simeq 0.2-0.3$fm 
 \cite{weise-vogl,CQ-size}, apart from
normalizations of the quark-gluon coupling constant.
The experimental value for the axial radius of the nucleon is
 $<r_A^2>^{1/2}\simeq 0.68$ fm
\cite{exp-r2A,hoferichter-etal}
and there are many estimations from lattice
$<r_A^2>^{1/2}\simeq 0.45- 0.50$ fm, for example in 
\cite{gupta-etal,alexandrou-etal} and references therein.

Similar behavior was found for the pseudoscalar 
quadratic radii presented in the next figure (\ref{fig:rqm-ps})
from expressions (\ref{r-ps-U},\ref{r-ps-U-tr}),
complete and truncated ones,
 as  functions of the quark effective mass $M^*$
for  the two gluon propagators.
The   non truncated  
expression  provides   negative  values and they are 
 presented with a sign minus.
One of them is 
divided by factor 10 for $D_I(k)$ to fit into a reasonable 
 scale of the figure.
To make possible a correct calculation with the previous figure
it was assumed $h_I=3$ and $h_{II}=1$.
The axial $<r^2>_A$ contribution  was found to be
smaller than the pseudoscalar $<r^2>_{ps}$ in all  cases.
This is related to the fact that  the pseudoscalar form factor 
normalization is larger than 
the axial form factor one. 
 At this level all the form factors reduce to 
only $F_1(K,Q)$ and $F_2(K,Q)$ and the truncated version $F_1^{tr}(K,Q)$.
However the difficulty in fixing the quark-gluon vertex and 
the overal momentum behavior of   
the quark and gluon propagators 
cannot be neglected.
When compared to the 
value
 $\sqrt{<r^2>_{CQ}} \simeq 0.2-0.3$fm
from \cite{weise-vogl,CQ-size}
the gluon propagator $D_I(k)$ provides larger values
for $<r^2>$
 and the gluon propagator
$D_{II}(k)$ again provides smaller values.
The reasons for the differences between
$<r^2>_{ps}$  and the truncated-$<r^2>_{ps}$  
 must be the same as  the ones responsible for the 
discrepancies in the axial radii from figure (\ref{fig:rqm-A}).
Besides that, it might be interesting, for the sake of comparison,
to compare with the
 scalar
radius of the lightest hadron,
 the pion, that
has been calculated, for example,  in lattice
with $<r^2>_s = 0.6$fm$^2$ \cite{rqm-pion-sca}.
The pion charge radius has estimations
for example  in lattice  $<r^2>=0.37$fm$^2$ \cite{charge-pi-rad}
and with SDE $<r^2>= 0.46-0.48$fm$^2$ \cite{tandy-maris-rpi},
whereas 
its experimental value  $<r^2>\simeq 0.45$fm$^2$ \cite{PDG,drechsel-walcher}.
The pion  scalar radius seems therefore to be
 larger than its charge radius
analogously to the fact that according to the present results
the pseudoscalar, and also scalar, radii are larger than the 
axial and vector radii.

\section{Summary and discussion}

 Pion-
constituent quark  momentum dependent  form factors 
were  investigated  from one loop background field method 
for the one non perturbative gluon exchange quark interaction
from the QCD effective action.
At this level, 
 the  pseudoscalar coupling only shows up for the 
usual pion field 
definition in terms of  unitary functions $U,U^\dagger$ but not for 
the Weinberg pion field.
Besides the usual pseudoscalar pion coupling,  
other  derivative  pion -scalar and 
pseudoscalar currents  form factors were also found in the leading order of the 
determinant expansion
in expressions (\ref{Gps-p-W},\ref{L-Q-pi-W}) and also 
(\ref{L-Q-pi-U},\ref{gps-k}).
Several of them have a reduced strength  with respect to the usual 
scalar and pseudoscalar form factors by a constant coefficient
of the order of $1/M^*$. 
By means of an integration by parts these terms might contribute
for the vector and axial channels.
All the (eleven) resulting form factors, pseudoscalar, scalar, vector and axial,
 were found to be written in terms of 
 only two momentum dependent functions $F_1(0,Q)$ and $F_2(0,Q)$ for 
zero external constituent  quark momentum,
with different coefficients.
A truncated  momentum 
dependence
of the quark kernel for $F_1(0,Q)$ was also considered such that 
the resulting form factors, $G_{A}^{W,tr} (0,Q)$ and 
$G_{ps}^{U,tr} (0,Q)$,  were shown to have a decreasing monotonic 
behavior more similar to the experimental results,
corresponding rather to the function  $F_2 (0,Q)$.
 The truncated expressions might in fact correspond to 
considering running momentum dependent effective sea 
quark mass from the
gap equation.
Besides that, these truncated expressions yield positive quadratic 
averaged radii.
Different values for the 
sea  quark effective mass $M^*$ were considered and 
it  mostly contributes for the overal normalization of the form factors.
 The first momentum dependent function presented was 
the constituent quark effective mass correction $M_3(Q)$.
Its momentum dependence is in excellent agreement with 
estimations from SDE calculations, except for its overall normalization
that appeared to be very large due to absence of
an UV  cutoff.
It is important to stress  that the mechanisms that give rise to 
the gap effective mass $M^*$ and to the mass $M_3(Q)$
are different.
However the behavior of constituent quark mass $M_3(Q)$  
is nearly independent of 
  the scalar condensate contribution for the
(constant) quark effective mass $M^*$.
At the level of the calculation presented, the axial and vector 
form factors 
are equal to each other for each of the pion field
definitions. 
The same chiral relation appeared for the scalar and pseudoscalar
form factors for the second pion definition.
 The axial and pseudoscalar form factors 
 were compared to  fittings
 of available  experimental data for pion nucleon
 form factors
by adjusting the values at zero momenta.
Results  showed  that the momentum dependence of 
 constituent quark
coupling to pions is not very different from the 
nucleon coupling to pions.
The larger 
difference between experimental (nucleon form factor) values
and the present form factors appear in the range
of $0.15 < Q < 1.4$GeV for $M^*=0.31$GeV.
This  might signal the need for improved momentum structure
of the quark and gluon kernels  but
it might also signal need to account for
effects from nucleon structure.
The pseudoscalar form factor has a larger strength 
than  the axial one, in agreement with expectations from 
phenomenology.
This conclusion remains valid if other
components for the axial form factor are included 
such as the coupling to light axial mesons,
as seen by comparing with results from Ref. \cite{PRD-2018a}
in which vector/axial mesons couplings to constituent quark 
had been investigated
by means of the same method employed in the present work.
A systematic and more general analysis will be presented elsewhere.
The pseudoscalar form factor  at the   timelike
point $Q^2=-m_\pi^2$, closer to  current
physical definitions of $g_{\pi N}$, 
 was obtained for the complete 
(or truncated)  expressions being smaller (or larger) than the
zero momentum $Q^2=0$ case.
 Different momentum dependent and independent 
ratios between the form factors were also presented.
Some of them simply show the resulting
chiral symmetry relations, eg. between vector and axial ones,
or between scalar and pseudoscalar ones.
The momentum dependence of the Goldberger Treiman 
relation (GTR) was also presented by considering 
the pseudoscalar and axial form factors for 
  spacelike   momenta and a qualitative agreement with 
calculations at the nucleon level as found.
  Finally the corresponding results for the 
 pseudoscalar and  contribution to the 
axial  constituent quark  averaged
quadratic radii were obtained
as functions of a constant  quark effective mass $M^*$
from the gap equation.
In particular resulting values for the axial/vector
quadratic radii are  somewhat smaller than
 estimations of the 
constituent quark  axial/vector radii from the coupling to light axial/vector mesons
obtained with the same method \cite{PRD-2018a}.
The    structureless pion limit might have  had effect on the estimations
but this structureless limit had also been considered for the vector/axial mesons.
In general the pseudoscalar quadratic radius 
 is larger than
the axial radius (from both couplings to pions and axial mesons)
 due to the corresponding form factors normalizations.
This becomes clear by noting all the quadratic radii and form factors depend
on only two momentum dependent functions.
The relevance of each of the constituent quark 
degree of freedom presented in this work and 
\cite{PRD-2018a} 
for nucleon structure and corresponding form factors
is to be investigated elsewhere.

\section*{Acknowledgments}

  The author thanks  short discussions 
with 
P. Bedaque, G.I. Krein  and  G. Eichmann, 
and  C.D. Roberts for kindly sending 
numerical results for the running quark mass
 from SDE from ref.
 \cite{craig-mass} plotted in Fig. (3).
The author participates to
 the project INCT-FNA, Proc. No. 464898/2014-5.


\begin{thebibliography}{00}



\bibitem{exp-r2A}
 K.L. Miller et al.,
{\it Study of the reaction $\nu_\mu d \to \mu^-  p p_s$},
 Phys. Rev. D 26, 537 (1982); 
T. Kitagaki et al.,
{\it High-energy quasielastic $\nu_\mu n \to \mu^- p$ 
scattering in deuterium},
 Phys. Rev. D 28, 436 (1983).


\bibitem{gaillard-savage} 
J-M. Gaillard, G. Sauvage, {\it Hyperon Beta Decays},
Ann. Rev. Nucl. Part. Sci. {\bf 34}, 351 (1984).



\bibitem{choi-etal-93}
S. Choi, {\it et al},
{\it Axial and Pseudoscalar Nucleon 
Form Factors from Low Energy Pion Electroproduction},
 Phys. Rev. Lett. {\bf 71}, 3927 (1993).




\bibitem{bardin-etal-1981}
G. Bardin {\it et al Measurement of the ortho para
 transition rate in the p mu p molecule and 
deduction of the pseudoscalar coupling constant $g_p^\mu$}, 
Phys. Lett. {\bf 104 B},  320  (1981).



\bibitem{andreev-etal-2007}
V.A. Andreev, {\it et al}., (MuCap Collaboration),
{\it 
Measurement of the Muon Capture 
Rate in Hydrogen Gas and Determination 
of the Proton’s Pseudoscalar Coupling $g_P$},
 Phys. Rev. Lett. {\bf 99}, 032002 (2007)








\bibitem{beise}
E.J.  Beise,{\it The Axial Form Factor of the Nucleon},
 Eur.Phys.J. A24S2, 43 (2005).

\bibitem{maris-craig}
P. Maris, C. D. Roberts,
{\it Dyson-Schwinger equations: A Tool for hadron physics},
Int. J. Mod. Phys. {\bf E 12},   297 (2003).
P. Tandy, 
{\it 
Hadron physics from the global color model of QCD},
 Prog.Part.Nucl.Phys. {\bf 39}, 117 (1997).

\bibitem{charge-pi-rad} 
J. van der Heide, J.H. Koch, E. Laermann,
{\it Pion structure from improved lattice QCD:
 form factor and charge radius at low masses},
Phys.Rev. D69, 094511 (2004).




\bibitem{drechsel-walcher} D. Drechsel, Th. Walcher,
{\it Hadron structure at low Q$^2$},
Rev.Mod.Phys. 80, 731 (2008).



\bibitem{yamazaki-etal} 
T. Yamazaki, {\it et al}, {\it Nucleon form factors with 2 +
 1 flavor dynamical domain-wall fermions},
Phys. Rev. {\bf D 79}, 114505 (2009).


\bibitem{constantinou} 
M. Constantinou, {\it Hadron Structure}, PoS(LATTICE2014)001.
arXiv:hep-latt: 1411.0078.



\bibitem{weise-vogl}
U. Vogl, W. Weise, 
{\it The Nambu and Jona-Lasinio model: Its implications for Hadrons and Nuclei},
 Progr. in Part. and Nucl. Phys. {\bf 27}  (1991) 195.

 


\bibitem{hoferichter-etal}
M. Hoferichter, C. Ditsche, B. Kubis, U.-G. Meissner,
{\it 
Dispersive analysis of the scalar form factor
of the nucleon},
JHEP 06, 063 (2012).





\bibitem{ramalho-tsushima-PRD94}
G. Ramalho, K. Tsushima, {\it Holographic estimate of the meson cloud contribution to 
nucleon axial form factor},
Phys. Rev. {\bf D 94}, 014001 (2016).




\bibitem{eichmann-fischer}
G. Eichmann, C. S. Fischer, 
{\it Nucleon axial and pseudoscalar form factors from the covariant Faddeev equation},
Eur. Phys. J. {\bf A 48},  9 (2012).




\bibitem{exp-bernard+E+meissner}
V. Bernard,  L. Elouadrihiri,
Ulf-G. Meissner,  {\it Axial structure of the nucleon},
J. Phys. G: Nucl. Part. Phys. {\bf 28} R1 (2002).



\bibitem{bratt-etal} J.D. Bratt, {\it et al}, {\it 
Nucleon structure from mixed action calculations using 2 + 1 flavors
of asqtad sea and domain wall valence fermions}, 
Phys. Rev. {\bf D 82}, 094502 (2010).
 




\bibitem{alexandrou-etal}
C. Alexandrou, {\it et al}, 
{\it Axial nucleon form factors from lattice QCD},
Phys. Rev. {\bf D 83}, 045010 (2011).
C. Alexandrou, {\it et al},  
{\it Nucleon axial form factors using lattice QCD simulations
with a physical value of the pion mass},
Phys. Rev. {\bf D 96}, 054507 (2017).

\bibitem{proton-radius}
A. Beyer et al.,
{\it 
 The Rydberg constant and proton size from atomic hydrogen}. 
Science. 358, 79  (2017).

\bibitem{PDG}
C. Patrignani {\it  et al} (Particle Data Group),
 Chin. Phys. C, 40,  (2016)  100001
and 2017 update.




\bibitem{weinberg-2010}
S. Weinberg, 
{\it Pions in Large 
N
 Quantum Chromodynamics},
Phys. Rev. Lett. {\bf 105}  (2010) 261601.   



\bibitem{constituent-1}
M. Lavelle, D. McMullan, {\it 
Constituent quarks from QCD},
 Phys. Rept. {\bf 279}, 1  (1997).
E. de Rafael,{\it The Constituent Chiral Quark Model revisited},
 Phys. Lett. {\bf B 703},   60  (2011).
 


\bibitem{constituent-2}
A. W. Thomas, Nucl. Phys. {\bf B} Proc. Suppl. {\bf 119}, 50 (2003).
R.D. Young, D.B. Leinweber, A.W. Thomas,
Progr. in Part. and Nucl. Phys. {\bf 50}, 399 (2003),
and references therein.



\bibitem{CQ-size}
Petronzio R., S. Simula and G. Ricco, 
{\it Possible evidence of extended objects inside the proton},
Phys. Rev. D 67, 094004 (2003) [Erratum-ibid. D 68, 099901
(2003)].



\bibitem{EPJA-2016}
 F.L. Braghin, {\it 
Quark and pion effective couplings from
polarization effects
}, 
Eur. Phys. Journ. {\bf A 52},  134  (2016).

\bibitem{EPJA-2018}
F.L. Braghin, {\it Low energy constituent quark and pion effective couplings in
a weak external magnetic field}, 
 Eur. Phys.  J. {\bf  A 54}, 45 (2018) .
ArXiv:1705.05926.

\bibitem{swanson-etal}
C.Downum, T.Barnes, J.R.Stone, E.S.Swanson,
{\it Nucleon-Meson Coupling Constants and Form Factors
in the Quark Model},
Phys.Lett. {\bf B 638}, 455 (2006).



\bibitem{PRD-2018a}
F.L. Braghin, {\it Light vector and axial mesons effective couplings to constituent quarks},
Phys. Rev. {\bf D 97}, 054025 (2018).

\bibitem{PRD-2018b}
F.L. Braghin,
{\it Constituent quark-light vector mesons effective couplings
in a weak background magnetic field},
 Phys. Rev. {\bf D 97},  014022 (2018).


\bibitem{proton-mass}
Y.B. Yang, J. Liang, Yu-J. Bi, Y. Chen, T. Draper, K.F. Liu, Z. Liu, 
{\it Proton Mass Decomposition from the QCD Energy Momentum Tensor}
Phys. Rev. Lett. {\bf 121}, 212001 (2018), and references therein.


\bibitem{cornwall} J. M. Cornwall,
{\it Entropy, confinement, and chiral symmetry breaking},
Phys. Rev. {\bf D 83}, 076001 (2011).

\bibitem{tandy-maris}
P. Maris, P.C. Tandy,
{\it Bethe-Salpeter study of vector meson masses and decay constants},
 Phys. Rev. C 60, 055214 (1999).





\bibitem{PRC1}
C.D. Roberts, R.T. Cahill, J. Praschifka,
{\it The Effective Action for the Goldstone Modes
in a Global Colour Symmetry Model of QCD},
 Ann. of Phys. {\bf 188}, 20  (1988).

\bibitem{ERV} D. Ebert, H. Reinhardt, M.K. Volkov,
{\it Effective hadron theory of QCD}, 
 Progr. Part. Nucl. Phys.
{\bf 33}, 1  (1994).

\bibitem{SD-rainbow}
D. Binosi, L. Chang, J. Papavassiliou, C.D. Roberts, 
{\it 
Bridging a gap between continuum-QCD and ab initio predictions of hadron observables},
Phys. Lett. {\bf B 742},  183  (2015) and references therein.




\bibitem{kondo}
K.-I. Kondo,
{\it Abelian-projected effective gauge theory of QCD 
with asymptotic freedom and quark confinement}
 Phys. Rev. {\bf D 57}, 7467 (1998) .

 \bibitem{higa}
 K. Higashijima, 
{\it Dynamical chiral-symmetry breaking},
Phys. Rev. D 29, 1228 (1984).


\bibitem{holdom}
B. Holdom, {\it 
Approaching low-energy QCD with a gauged, nonlocal, constituent-quark model},
Phys. Rev. {\bf D 45}, 2534 (1992).



\bibitem{wang-etal}
Q. Wang, Y.-P. Kuang, X.-L. Wang, and M. Xiao, Phys.
Rev. D 61, 054011 (2000); K. Ren, H.-F. Fu, and Q. Wang,
Phys. Rev. D 95, 074012 (2017).




\bibitem{PRD-2016}
F.L. Braghin, 
{\it SU(2) low energy quark effective couplings in weak external magnetic field} ,
Phys. Rev. {\bf D 94} (2016) 074030.


\bibitem{background}
L.F. Abbott, 
Acta Phys. Pol. {\bf B 13}, 33 (1982).


\bibitem{SWbook}
S. Weinberg, {\it The Quantum Theory of Fields} Vol. II, Cambridge, 
(1996).



\bibitem{craig-bloch-schimidt}
J.C. Bloch, C.D. Roberts, S.M. Schmidt,
{\it Selected nucleon form factors and a composite scalar diquark},
Phys. Rev. {\bf C 61}, 065207 (2000).



\bibitem{craig-mass}
C. Chen, B. El-Bennich, C. D. Roberts, S. M. Schmidt, 
J. Segovia, S. Wan,
{\it Structure of the nucleon's low-lying excitations},
 Phys. Rev. D 97,  034016 (2018),  arXiv:1711.03142 [nucl-th],


\bibitem{sasaki-yamazaki}
S. Sasaki, T. Yamazaki,
{\it Nucleon form factors from quenched lattice QCD
 with domain wall fermions}, 
Phys. Rev.  {\bf D 78},  014510 (2008).

 
\bibitem{gupta-etal}
R. Gupta, Y.-C. Jang, H.-W. Lin, B. Yoon, T. Bhattacharya,
{\it Axial-vector form factors of the nucleon from lattice QCD},
Phys. Rev. D96 (2017) 114503. arXiv:1705.06834, 
Y.-C. Jang, T. Bhattacharya, R. Gupta, H.-W. Lin, B. Yoon,
{\it Nucleon Axial and Electromagnetic Form Factors},
 EPJ Web
Conf. 175 (2018) 06033. arXiv:1801.01635, doi:10.1051/
epjconf/201817506033.





\bibitem{rqm-pion-sca}  V. Gulpers, G. von Hippel,  H. Wittig,
{\it The scalar radius of the pion from lattice QCD in the continuum limit},
Eur. Phys. J. A 51, 158 (2015).

\bibitem{tandy-maris-rpi}
P. Maris, P.C. Tandy, {\it 
The Quark-Photon Vertex and the Pion Charge Radius},
arXiv:nucl-th/9910033.  

\end{thebibliography}
\end{document}